\documentclass[pre,twocolumn,showpacs]{revtex4-1}
\usepackage{amsmath,amssymb}
\usepackage{graphicx}
\usepackage{epsfig}

\begin{document}
\title{Spin-glass splitting in the quantum Ghatak-Sherrington model}
\author{Do-Hyun Kim}

\affiliation{Jesuit Community, Sogang University, 35 Baekbeom-ro, Mapo-gu, Seoul 121-742, Korea}

\begin{abstract}
We propose an expanded spin-glass model, called the quantum Ghatak-Sherrington model, which considers
spin-1 quantum spin operators in a crystal field and in a transverse field. The analytic solutions and phase
diagrams of this model are obtained by using the one-step replica symmetry-breaking ansatz under the static
approximation. Our results represent the splitting within one spin-glass (SG) phase depending on the values
of crystal and transverse fields. The two separated SG phases, characterized by a density of filled states, show
certain differences in their shapes and phase boundaries. Such SG splitting becomes more distinctive when the
degeneracy of the empty states of spins is larger than one of their filled states.
\end{abstract}


\maketitle

\section{Introduction}

A spin glass (SG) is a complex system characterized by
both quenched randomness and frustration, which lead to the
irreversible freezing of spins to states without the long-range
spatial order below the glass transition temperature ($T_{g}$) \cite{Nishimori}.
Theoretical approaches for understanding SG transitions are
generally concerned with the study of mean-field level calculations performed using infinite-range interaction models,
of which the Sherrington-Kirkpatrick (SK) model \cite{SK} is a
prototype. Some infinite-range interaction SG models have
recently sparked interest in relation to the so-called inverse
transitions.

Since Tammann's hypothesis \cite{Tammann} a century ago, there has
been substantial interest in a different class of phase transitions
known as inverse transitions (melting or freezing). In these
phase transitions, an ordered phase is more entropic than a
disordered one, whereby the ordered phase may appear at
a higher temperature than the disordered one. Such inverse
transitions have already been observed experimentally in
physical systems such as of liquid crystals \cite{Cladis77},
polymers \cite{Rastogi99,Chevillard97}, high-$T_{c}$ superconductors \cite{Avraham01},
magnetic thin films \cite{Portmann03}, and organic monolayers \cite{Scholl10}. Meanwhile, from a theoretical
point of view, there have been various attempts to identify
a suitable model for inverse transitions. Spin-glass models
have been suggested to be candidates for inverse freezing,
wherein the SG phase becomes one with higher entropy. The
Ghatak-Sherrington (GS) model \cite{GS,Costa94} is a spin-1 spin-glass
model with a crystal field and it is especially well known as a
prototypical SG model for inverse freezing \cite{Schupper04,Crisanti05,Leuzzi06,Magalhaes08,Morais12}.

In ordinary SG systems, in general, the second-order
phase transition from paramagnetic (PM) to SG occurs as
temperature is decreased. However, according to Crisanti and
Leuzzi \cite{Crisanti05,Leuzzi06}, there seems to be a second reentrance as well as
inverse freezing in the GS model. (See FIG. 2 in Refs. \cite{Crisanti05,Leuzzi06}.)
This implies that phase transitions are likely when the phase
is varied successively in the order PM $\rightarrow$ SG $\rightarrow$ PM $\rightarrow$ SG
as the temperature is reduced. In other words, there seems to
exist two different SGs, i.e., a SG in the higher-temperature
region [higher-temperature spin glass (HTSG)] and a SG in
the lower-temperature region [lower-temperature spin glass
(LTSG)]. The aim of this paper is to investigate the theoretical
validity for the existence of such separated SGs using a simple
GS-like model.

For this purpose, we study a quantum version of the GS
model by adding a transverse tunneling field, similar to the
manner in which the quantum version of the SK model
has been studied by considering quantum tunneling with a
transverse field \cite{Chakrabarti96,Kim02}. We expect the quantum GS model
to clarify the changes in the existence and features of the
two SGs with respect to the transverse field. Herein we use
one-step replica symmetry breaking (1RSB) for theoretical
investigations instead of the replica symmetry (RS) \cite{Schupper04,Crisanti05,Leuzzi06}
and the full replica symmetry breaking (FRSB) \cite{Crisanti05,Leuzzi06}.
We select the 1RSB because it provides more physically
meaningful results than RS does and numerical values of order
parameters more easily than FRSB does. Although 1RSB
is approximated with respect to the exact FRSB ansatz, it
is a good approximation around transition lines because at
criticality the thermodynamics is not very sensitive to the
ansatz chosen, as shown in Refs. \cite{Crisanti05,Leuzzi06}.

\section{Model}

The Hamiltonian of the quantum GS model is
\begin{eqnarray}
\mathcal{H} = - \sum_{(i,j)} J_{ij} S_{iz} S_{jz} + D \sum_{i} S_{iz}^{2} - \Gamma \sum_{i} S_{ix}
\end{eqnarray}
where ($i,j$) means all the distinct pairs of spins with the total number $N$,
$J_{ij}$ are quenched random exchange interaction variables, $D$ is the crystal field, and $\Gamma$
is the transverse tunneling field. The spin-1 quantum spin operators $S_{z}$ and $S_{x}$ are defined by
\begin{eqnarray}
S_{z}=\left(
\begin{array}{lll}
1~~~0~~~0\\
0~~~0~~~0\\
0~~~0~-1
\end{array}\right)~~\textrm{and}~~
S_{x}=\frac{1}{\sqrt{2}}\left(
\begin{array}{lll}
0~~~1~~~0\\
1~~~0~~~1\\
0~~~1~~~0
\end{array}\right),
\end{eqnarray}
respectively. The distribution of $J_{ij}$ is taken to be Gaussian with a
mean zero and a variance of $1/N$. When $k$ and $l$ are the degeneracy of the filled or interacting states
of $S_{z}$ and of the empty or noninteracting states of $S_{z}$,
respectively, we can define the relative degeneracy of the
filled states as $r \equiv k/l$ \cite{Schupper04,Crisanti05,Leuzzi06}.

By the imaginary-time formalism \cite{Bray80}, the partition
function of the system can be written as
\begin{eqnarray}
Z &=& \textrm{Tr}~\exp \Big[ \beta \Gamma \sum_{i}^{N}
S_{ix} \Big] \mathcal{T} \exp \Big[\int_{0}^{\beta} d\tau \nonumber\\
&& \Big\{ \sum_{ij}^{N} J_{ij}S_{iz}(\tau)S_{jz}(\tau) - D \sum_{i}^{N} (S_{iz}(\tau))^{2} \Big\}\Big]
\end{eqnarray}
where $\tau$ is the imaginary time, $\mathcal{T}$ is the time-ordering operator,
$S_{iz}(\tau)$ are the operators under the
interaction representation introduced in the quantum physics,
[i.e., $S_{iz}(\tau) \equiv \exp(\mathcal{H}_{0}\tau)
S_{iz} \exp(-\mathcal{H}_{0}\tau)$ where $\mathcal{H}_{0} =
-\Gamma \sum_{i}^{N} S_{ix}$] and $\beta = 1/T$ (where
$k_{B}\equiv 1$ for simplicity). For
this model, the free energy is calculated as $-\beta F
\equiv [\ln Z]_{J} = \int \prod_{i,j}^{N} dJ_{ij} P(J_{ij})
\ln Z (\{J_{ij}\})$, where $[~~]_{J}$
indicates an average over the quenched disorder of $J_{ij}$. For the quenched random system the
free energy can be evaluated using the replica method
$\ln Z = \lim_{n \to 0} (1/n)[Z^{n}-1]$.

By averaging $Z^{n}$ over $P(J_{ij})$, rearranging
terms, and taking the method of steepest descent in the
thermodynamic limit ($N \rightarrow \infty$), the
intensive free energy $f \equiv \lim_{N \to \infty} F/N$ can be written as
\begin{widetext}
\begin{eqnarray}
\beta f = \lim_{n \to 0}\frac{1}{n} \bigg\{
\frac{1}{4} \int_{0}^{\beta} d\tau
\int_{0}^{\beta} d\tau' \Big[\sum_{(\alpha \beta)}^{n}
\big(Q^{\alpha \beta}(\tau, \tau')\big)^{2} + \sum_{\alpha}^{n}
\big(R^{\alpha \alpha}(\tau, \tau')\big)^{2}\Big] - \ln \textrm{Tr} \exp
(\tilde{\mathcal{H}}) \bigg\}
\end{eqnarray}
with the effective Hamiltonian
\begin{eqnarray}
\exp (\tilde{\mathcal{H}}) \equiv \exp \Big[ \beta \Gamma
\sum_{\alpha}^{n} S_{x}^{\alpha} \Big] ~\mathcal{T} \exp \bigg\{
\int_{0}^{\beta} d\tau \int_{0}^{\beta} d\tau'
\Big[ \frac{1}{2} \sum_{(\alpha \beta)}^{n}
Q^{\alpha \beta}(\tau, \tau') S_{z}^{\alpha}(\tau)S_{z}^{\beta}(\tau')  \nonumber\\
+ \frac{1}{2} \sum_{\alpha}^{n} R^{\alpha \alpha}(\tau, \tau')
S_{z}^{\alpha}(\tau)S_{z}^{\alpha}(\tau') \Big]
- D \int_{0}^{\beta} d\tau \sum_{\alpha}^{n} (S_{z}^{\alpha}(\tau))^{2}
\bigg\}
\end{eqnarray}
\end{widetext}
where $(\alpha \beta)$ denotes a summation over replica indices
$\alpha$ and $\beta (\neq \alpha)$ running from 1 to $n$, and the
trace $\textrm{Tr}$ is over $n$ replicas at a single spin
site. Here two order parameters are
introduced: the spin-glass order parameter
$Q^{\alpha \beta}(\tau, \tau') \equiv \langle \mathcal{T}
S_{z}^{\alpha}(\tau)S_{z}^{\beta}(\tau') \rangle$ and the spin self-interaction $R^{\alpha \alpha}(\tau, \tau') \equiv \langle \mathcal{T}
S_{z}^{\alpha}(\tau)S_{z}^{\alpha}(\tau') \rangle$, where
$\langle A \rangle \equiv \textrm{Tr}~ [A~e^{\tilde{\mathcal{H}}}]/\textrm{Tr}~ e^{\tilde{\mathcal{H}}}$.

We take the static approximation
\cite{Bray80} by $Q^{\alpha \beta}(\tau,
\tau')=Q^{\alpha \beta}$ and $R^{\alpha \alpha}(\tau,
\tau')=R^{\alpha \alpha}$. Then the free energy $f$ is given by
\begin{widetext}
\begin{eqnarray}
\beta f = \lim_{n \to 0} \frac{1}{n}
\Big[\frac{1}{4} \beta^{2}\Big\{\sum_{(\alpha \beta)}^{n}
(Q^{\alpha \beta})^{2} + \sum_{\alpha}^{n} (R^{\alpha
\alpha})^{2}\Big\} - \ln \textrm{Tr} \exp (\tilde{\mathcal{H}}') \Big]~~
\end{eqnarray}
with the effective Hamiltonian
\begin{eqnarray}
\tilde{\mathcal{H}}' \equiv \frac{1}{2} \beta^{2}
\sum_{(\alpha \beta)}^{n} Q^{\alpha \beta}
S_{z}^{\alpha}S_{z}^{\beta} + \sum_{\alpha}^{n} \Big( \frac{1}{2} \beta^{2} R^{\alpha
\alpha} - \beta D \Big) (S_{z}^{\alpha})^{2} + \beta \Gamma
\sum_{\alpha}^{n} S_{x}^{\alpha}
\end{eqnarray}
\end{widetext}

Next, we use Parisi's 1RSB scheme as in the case of the SK model
\cite{Parisi}: for the $n \times n$ matrix $\{Q^{\alpha \beta}\}$ in
the replica spin space, the $n$ replicas of $\{Q^{\alpha \beta}\}$
are divided into $n/m$ groups of $m$ replicas, assuming that $n$
must be a multiple of $m$, so that $\{Q^{\alpha \beta}\}$ consists
of $n/m$ diagonal matrices of $m \times m$ elements each (in which all the diagonal
elements are zero and off-diagonal elements are $Q_{1}$) and $n/m \times (n/m-1)$ matrices of $m \times m$ elements (in which all the elements are
$Q_{0}$). Then the free energy obtained by the 1RSB ansatz is given as follows:
\begin{widetext}
\begin{eqnarray}
\beta f_{_{1RSB}} = \frac{1}{4}
\beta^{2}\{R^{2}-Q_{1}^{2} + m(Q_{1}^{2}-Q_{0}^{2})\}
-\frac{1}{m} \int \mathcal{D}z ~\ln \bigg[ \int \mathcal{D}y~ \Big[1 + 2 re^{\gamma} \cosh
\big(\beta \sqrt{H(z,y)^{2}+ \Gamma^{2}}\big) \Big]^{m} \bigg]
\end{eqnarray}
where $\int \mathcal{D}z(y) ~ \cdots \equiv \frac{1}{\sqrt{2 \pi}}
\int_{-\infty}^{\infty} dz(y) ~ e^{-[z(y)]^{2}/2} \cdots $, $\gamma \equiv \frac{1}{2}\beta^{2} (R-Q_{1}) - \beta D$, and  $H(z,y) \equiv \sqrt{Q_{0}}~z + \sqrt{Q_{1}-Q_{0}}~y$.
The self-consistent equations for $m$, $Q_{0}$, $Q_{1}$, and $R$ are obtained by the extremal condition of
$f_{_{1RSB}}$:
\begin{eqnarray}
&&\frac{1}{4} \beta^{2}m^{2}(Q_{1}^{2}-Q_{0}^{2})= -\int
\mathcal{D}z ~\ln \bigg[ \int \mathcal{D}y~ \Big[1 + 2 re^{\gamma} \cosh
\big(\beta \sqrt{H(z,y)^{2}+ \Gamma^{2}}\big) \Big]^{m} \bigg] \nonumber\\
&&~~~+ m\int \mathcal{D}z~ \frac{\int \mathcal{D}y~ \Big[1 + 2 re^{\gamma} \cosh
\big(\beta \sqrt{H(z,y)^{2}+ \Gamma^{2}}\big) \Big]^{m} \ln \Big[1 + 2 re^{\gamma} \cosh
\big(\beta \sqrt{H(z,y)^{2}+ \Gamma^{2}}\big) \Big]}{\int
\mathcal{D}y~ \Big[1 + 2 re^{\gamma} \cosh
\big(\beta \sqrt{H(z,y)^{2}+ \Gamma^{2}}\big) \Big]^{m}}\\
&&R=\int \mathcal{D}z~ \frac{\int \mathcal{D}y~ \Big[1 + 2 re^{\gamma} \cosh
\big(\beta \sqrt{H(z,y)^{2}+ \Gamma^{2}}\big) \Big]^{m} \frac{H(z,y)^{2}}{H(z,y)^{2}+ \Gamma^{2}}  \bigg[\frac{2 re^{\gamma}
\cosh \big(\beta \sqrt{H(z,y)^{2}+ \Gamma^{2}}\big)}{1+2 re^{\gamma} \cosh \big(\beta \sqrt{H(z,y)^{2}+ \Gamma^{2}}\big)}\bigg]}{\int
\mathcal{D}y~ \Big[1 + 2 re^{\gamma} \cosh
\big(\beta \sqrt{H(z,y)^{2}+ \Gamma^{2}}\big) \Big]^{m} }\\
&&Q_{0}=\int \mathcal{D}z~ \Bigg[ \frac{\int \mathcal{D}y~
\Big[1 + 2 re^{\gamma} \cosh
\big(\beta \sqrt{H(z,y)^{2}+ \Gamma^{2}}\big) \Big]^{m} \frac{H(z,y)}{\sqrt{H(z,y)^{2}+ \Gamma^{2}}} \bigg[\frac{2 re^{\gamma}
\sinh \big(\beta \sqrt{H(z,y)^{2}+ \Gamma^{2}}\big)}{1+2 re^{\gamma} \cosh \big(\beta \sqrt{H(z,y)^{2}+ \Gamma^{2}}\big)}\bigg]}{\int
\mathcal{D}y~ \Big[1 + 2 re^{\gamma} \cosh
\big(\beta \sqrt{H(z,y)^{2}+ \Gamma^{2}}\big) \Big]^{m} } \Bigg]^{2}\\
&&Q_{1}=\int \mathcal{D}z~ \frac{\int \mathcal{D}y~ \Big[1 + 2 re^{\gamma} \cosh
\big(\beta \sqrt{H(z,y)^{2}+ \Gamma^{2}}\big) \Big]^{m} \frac{H(z,y)^{2}}{H(z,y)^{2}+ \Gamma^{2}} \bigg[\frac{2 re^{\gamma}
\sinh \big(\beta \sqrt{H(z,y)^{2}+ \Gamma^{2}}\big)}{1+2 re^{\gamma} \cosh \big(\beta \sqrt{H(z,y)^{2}+ \Gamma^{2}}\big)}\bigg]^{2}}{\int
\mathcal{D}y~ \Big[1 + 2 re^{\gamma} \cosh
\big(\beta \sqrt{H(z,y)^{2}+ \Gamma^{2}}\big) \Big]^{m} }
\end{eqnarray}
\end{widetext}
We can complete phase diagrams of the present model from these equations.

\section{Results}

First, let us consider the $r=1$ case in order to check whether the result of Crisanti and Leuzzi \cite{Crisanti05,Leuzzi06} is correct. The graphs in Fig. 1 show the $T-D$ phase diagrams obtained for specific $\Gamma$ values. As shown in Fig. 1(a), the $T-D$ phase diagram of the $\Gamma=0.0$ case (GS model) at $r=1$ is nearly the same as that of the model used by Crisanti and Leuzzi \cite{Crisanti05,Leuzzi06}. The locations of the first-order phase boundary and tricritical point (TCP), i.e., the cross-point between first- and second-order phase boundaries, were determined by the same criteria proposed in Ref. \cite{Costa94}. The TCP of Fig. 1(a) is located at (0.962, 0.333), as analytically obtained in Ref. \cite{Costa94}. In the region $0.0 \leq D < 0.879$, the second-order phase transition from PM to SG occurs as the temperature is decreased, which is generally observed in ordinary SG systems. However, in the region $0.879 \leq D < 0.9$, successive phase transitions occur for which the phase is varied in the order PM $\stackrel{\mathrm{2nd}}{\longrightarrow}$ HTSG $\stackrel{\mathrm{1st}}{\longrightarrow}$ PM $\stackrel{\mathrm{1st}}{\longrightarrow}$ LTSG, as the temperature is reduced. This result shows clearly the second reentrance that Crisanti and Leuzzi referred to previously \cite{Crisanti05,Leuzzi06}. In the region $0.9 \leq D \leq 0.962$, inverse freezing is shown through the phase transitions in the order PM $\stackrel{\mathrm{2nd}}{\longrightarrow}$ SG $\stackrel{\mathrm{1st}}{\longrightarrow}$ PM, as the temperature is decreased. Therefore, we have verified that inverse freezing, which many investigators of the GS model have focused upon, occurs only in a narrow region.

Figure. 1(b) shows the $T-D$ phase diagrams for several values of $\Gamma$, including the result of the $\Gamma=0.0$ case (GS model). As $\Gamma$ is gradually increased, the glass transition temperatures decrease. In the range $0.0 \leq D \lesssim 0.7$, only the second-order phase transition from PM to SG occurs as the temperature is reduced, and the glass transition temperatures decrease as $\Gamma$ is increased. However, when $D$ is larger than 0.7, the first-order phase transitions occur and the position of each TCP depends on each $\Gamma$ value. The shapes of the phase boundaries in this range are rather complex, as can be checked in Fig. 2(b).

\begin{figure}
\includegraphics[width=0.5\textwidth]{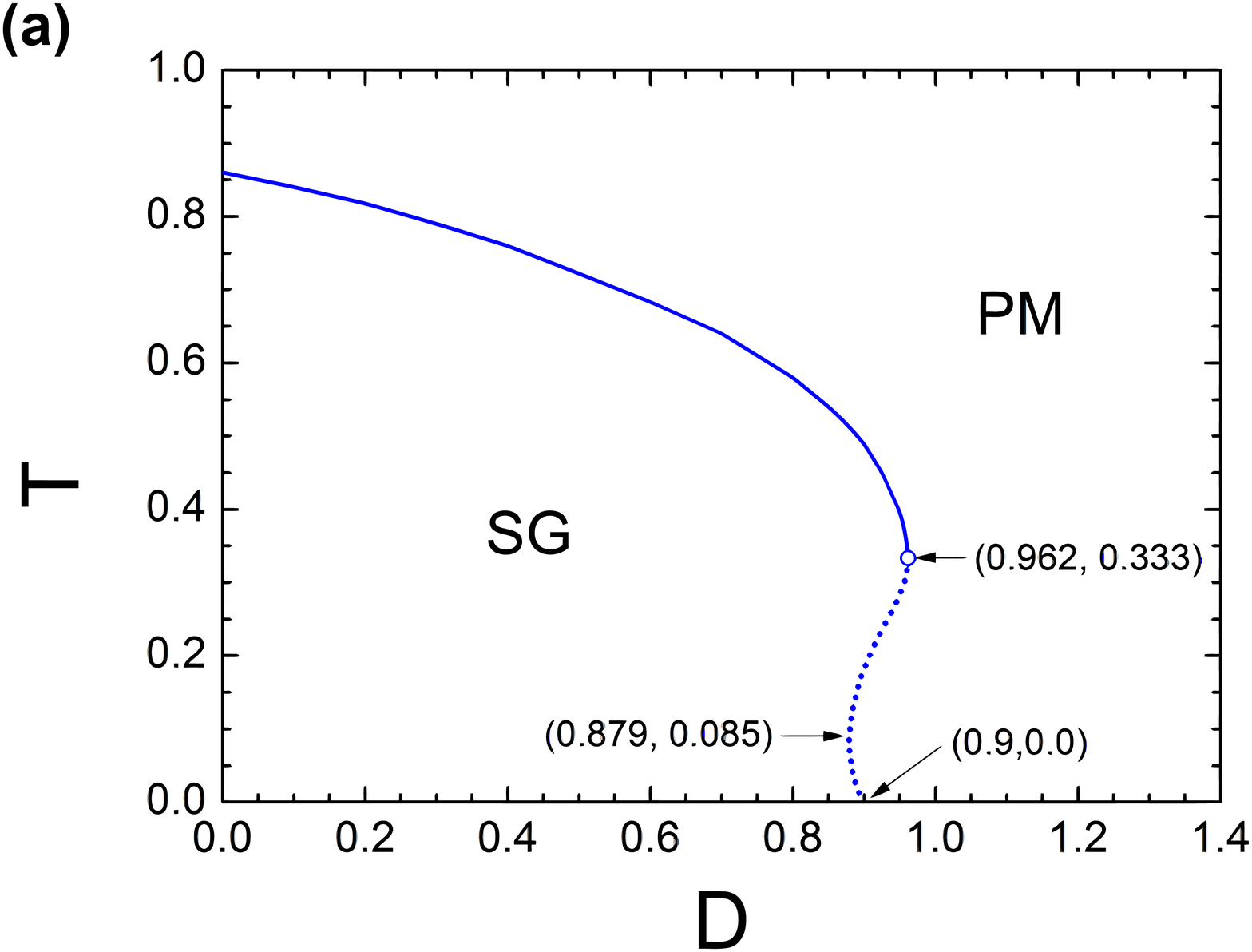}
\includegraphics[width=0.5\textwidth]{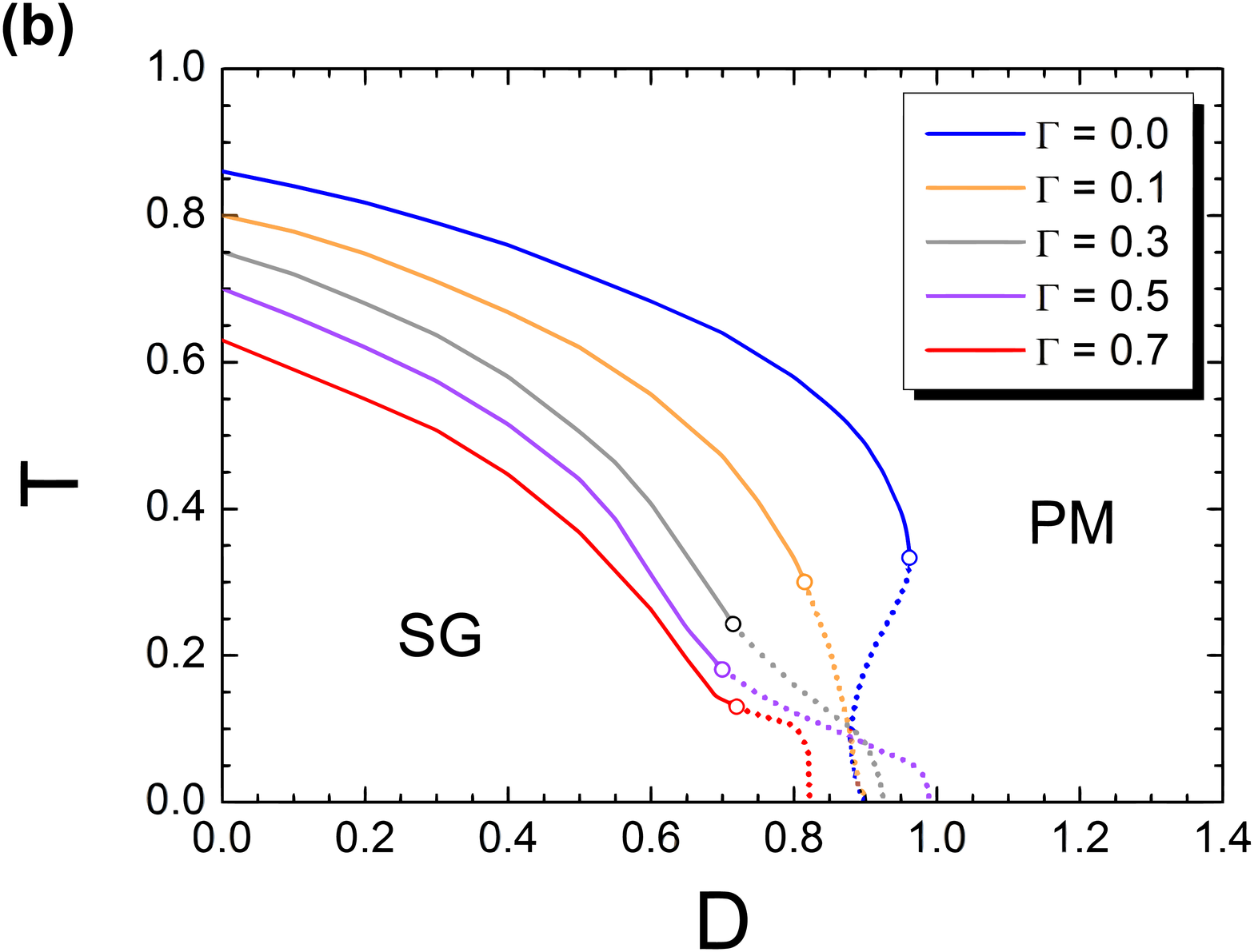}
\caption{(Color online) (a) $T-D$ phase diagram for the $\Gamma=0.0$ case (GS model) and
(b)	$T-D$ phase diagrams for several values of $\Gamma$. The solid-line (dotted-line) part of each phase boundary indicates the second-order (first-order) phase transition and each circle between the two kinds of lines denotes a TCP.}\label{1}
\end{figure}

\begin{figure}
\includegraphics[width=0.42\textwidth]{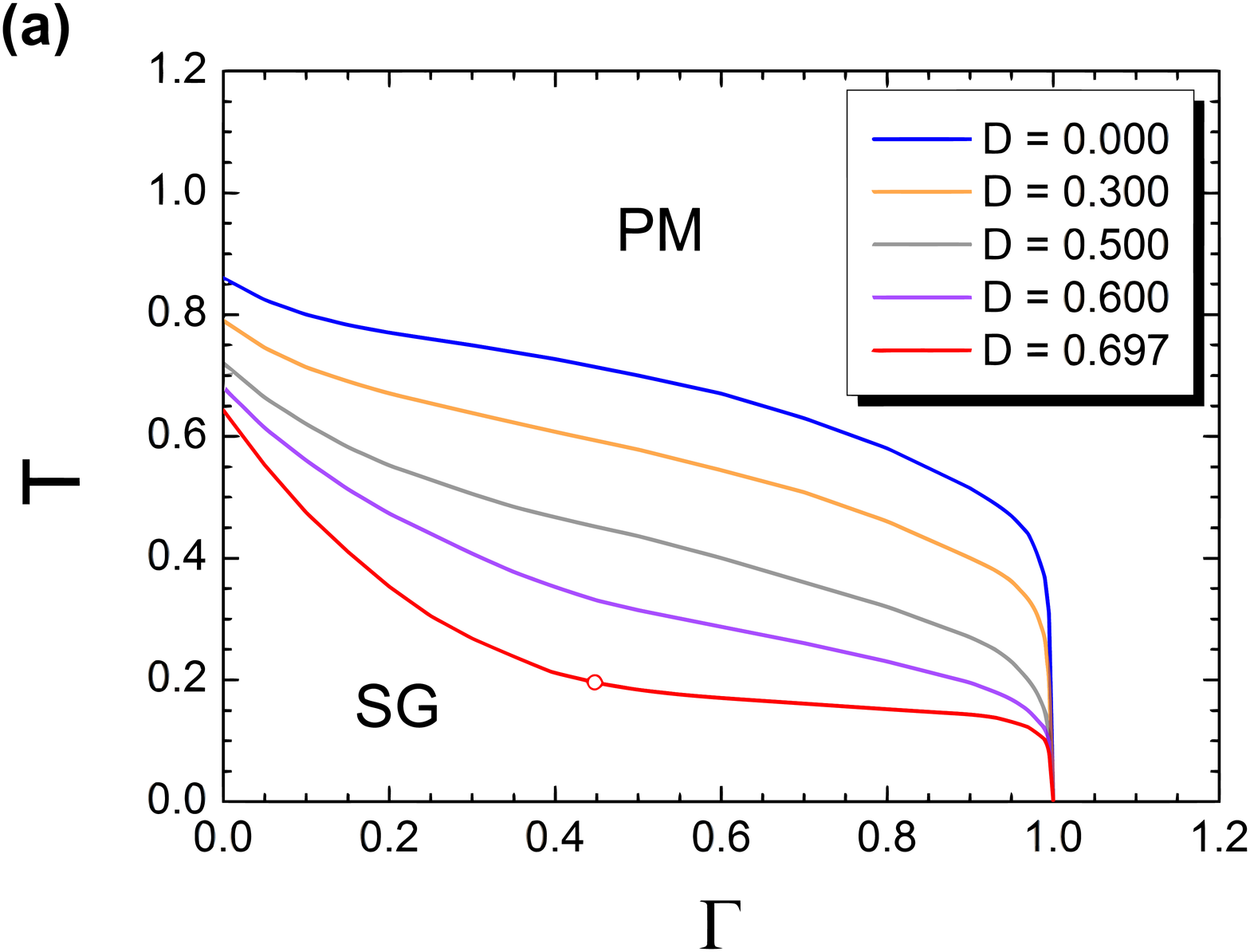}
\includegraphics[width=0.42\textwidth]{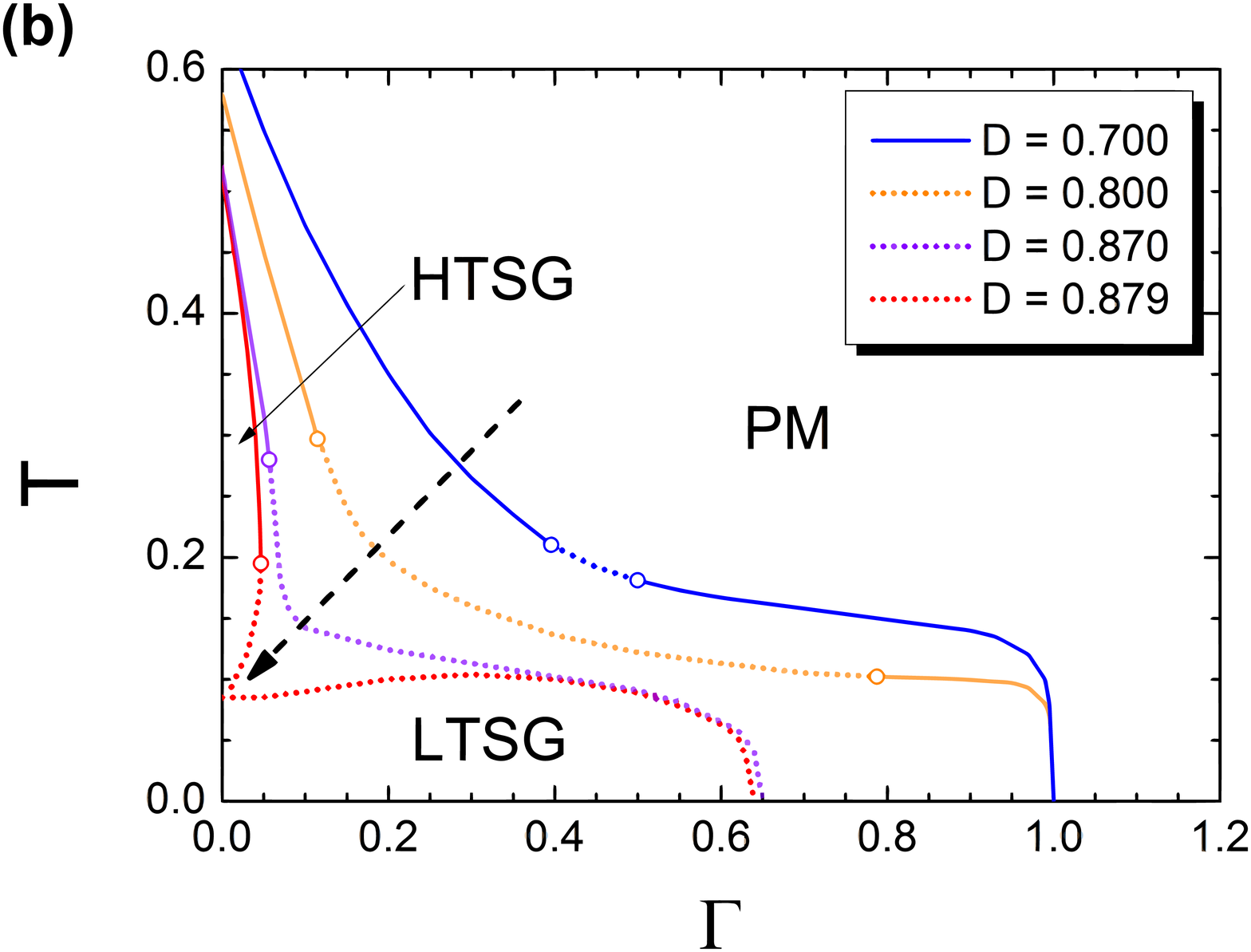}
\includegraphics[width=0.42\textwidth]{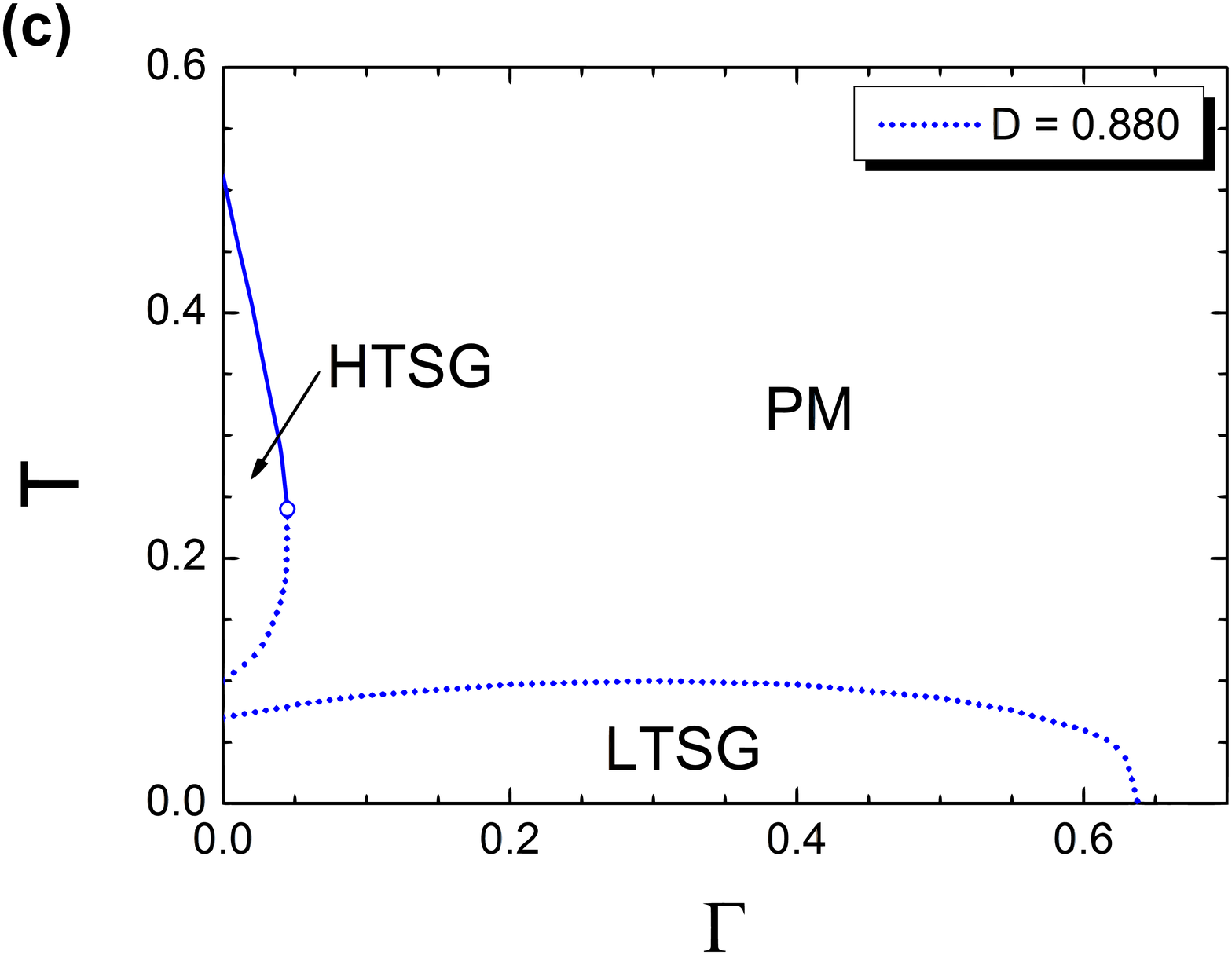}
\includegraphics[width=0.42\textwidth]{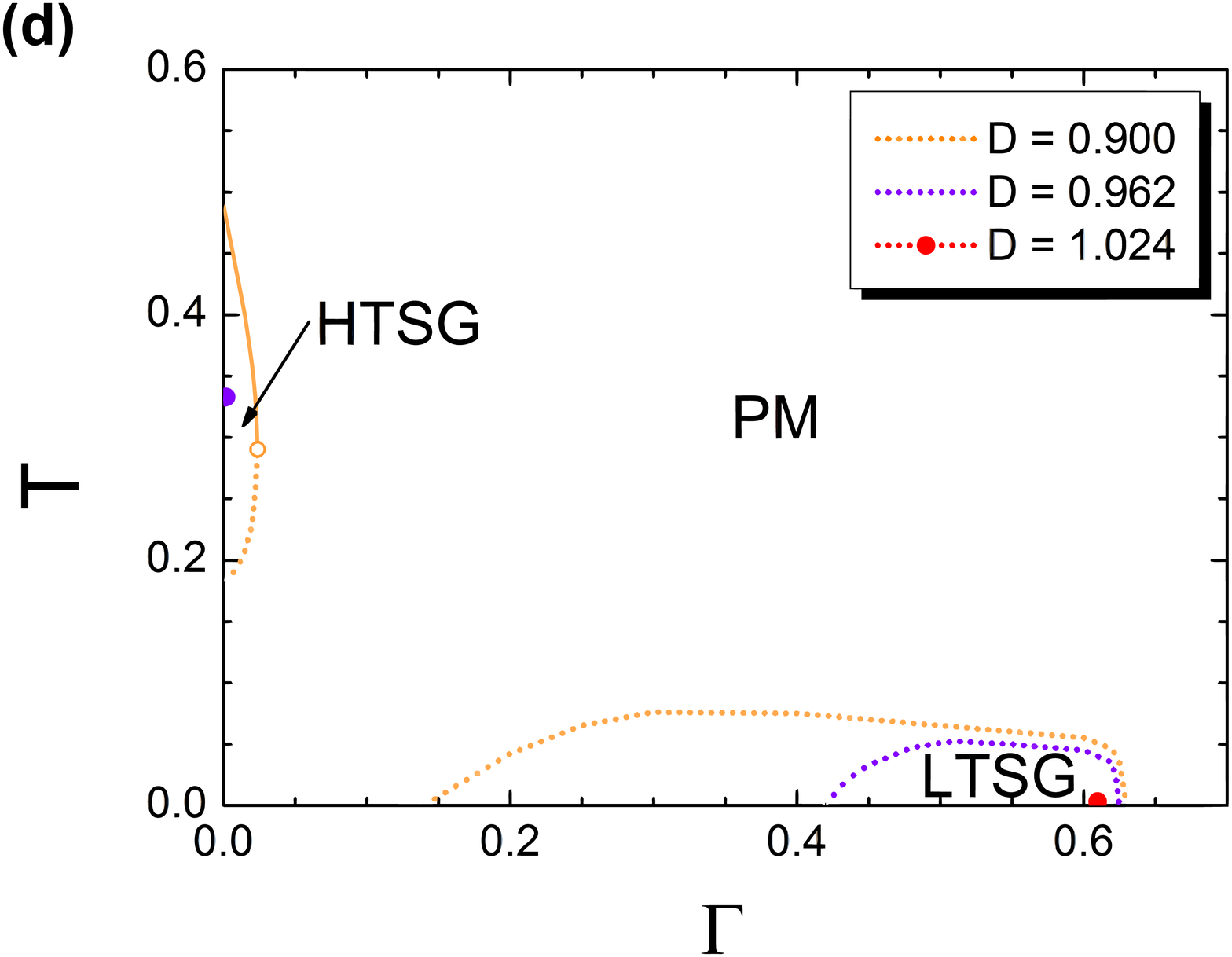}
\caption{(Color online) The $T-\Gamma$ phase diagrams for several values of $D$ between (a) 0.0 and 0.697 and (b) 0.697 and 0.879.
As $D$ increases gradually, the phase boundary is kinked in the direction of the dashed arrow of the figure.
(c) The case of $D=0.88$. When $D$ is larger than 0.879, the phase boundary becomes split. (d) Three cases with $D$ larger than 0.88. }\label{2}
\end{figure}

The graphs of Fig. 2 show the $T-\Gamma$ phase diagrams obtained for specific $D$ values. Figure 2(a) represents the temperature-dependent variations in the phase boundaries, which are obtained for $D$ between 0.0 and 0.697. In the Ising spin-glass model with a transverse field, the glass transition temperature at $\Gamma=0.0$ is 1.0 \cite{Kim02}, whereas in our model the transition temperature at $\Gamma=0.0$ is 0.86. The difference between the two values can be attributed to the fact that our model includes the eigenvalues of $S_{z} = 0$ as well as $S_{z} = 1$ and $-1$. When $r$ is increased, the glass transition temperature gradually increases to 1.0. As expected, the phase boundary is shifted to a lower temperature with the increase in $D$. According to our detailed numerical calculation, the first-order phase transition first arises at $D=0.697$, where the TCP is located at $(\Gamma, T) = (0.448, 0.196)$.

When $D$ is larger than 0.697, the shift becomes more complex, as shown in Fig. 2(b). As $D$ is larger than 0.697, one TCP is separated into two new TCPs and a first-order phase transition lies between these two TCPs \cite{comment}. As $D$ is gradually increased, the phase boundary is kinked in the direction of the dashed arrow of Fig. 2(b) and the region of the first-order phase transition simultaneously broadens. As $D$ increases further, one of the TCPs collapses with the $\Gamma$ axis.
When $D$ becomes 0.879, the phase boundary starts to split. The second reentrance of the GS model [Fig. 1(a)] is a zero-$\Gamma$ case reflecting this splitting of the phase boundary. The two SG phases generated by the splitting are the HTSG and the LTSG. The HTSG is inside the extremely narrow region of $\Gamma$ and surrounded by the $T$ axis, the second-order phase boundary, one TCP, and the first-order phase boundary. However, the LTSG is spread along the $\Gamma$ axis and is surrounded only by the axis and the first-order phase boundary.

In the case of $D=0.88$ of Fig. 2(c), two types of SGs (HTSG and LTSG) exist between $0.0 \leq \Gamma \leq 0.045$. However, for values of $\Gamma$ greater than 0.045, only one type of SG (LTSG) exists under the PM phase.

The case of $D=0.9$ in Fig. 2(d) is characterized by the clear occurrence of inverse freezing in the extremely narrow region of $0.0 \leq \Gamma \leq 0.024$. However, for $0.024 < \Gamma < 0.14$, there is no other phase except the PM phase at any temperature. For the $0.14 \leq \Gamma \leq 0.63$ region, the LTSG exists under the PM phase.
When $D$ reaches the value of 0.962, the HTSG converges to one point $(\Gamma, T) = (0.0, 0.333)$, which is the TCP of the GS model. Therefore, as $D$ increases, one TCP corresponding to the $D$ value greater than 0.697 gradually shifts to the TCP of the GS model, and the area of the HTSG reduced throughout this process, until the HTSG converges to the TCP of the GS model. During the same process, the area of the LTSG also decreases gradually. When $D$ reaches the value of 1.024, the LTSG converges to a point $(\Gamma, T) = (0.61, 0.0)$. When $D$ is larger than 1.024, no SG phase exists for any temperature or $\Gamma$ field. The appearance and disappearance of the HTSG and LTSG thus depend on the value of $D$.

Note that the two SG phases (HTSG and LTSG) originate from the $D$ field,
irrespective of the $\Gamma$ field. As shown in Fig. 1(a), in the region $0.879 \leq D < 0.9$,
the two SG phases occur even when $\Gamma = 0$. The role of the $\Gamma$ field is to lower
the glass transition temperature through quantum tunneling in proportion to the $\Gamma$ value,
as already checked in Refs. \cite{Chakrabarti96,Kim02}. In particular, in our model,
the $\Gamma$ field plays a role in the sudden lowering of the second-order transition temperature of the SG (at $D < 0.879$) or HTSG (at $0.879 \leq D < 0.962$). Thus even a small value of the $\Gamma$ field (about 0.05) makes the HTSG disappear in the region $0.879 \leq D < 0.962$.

\begin{figure}
\includegraphics[width=0.42\textwidth]{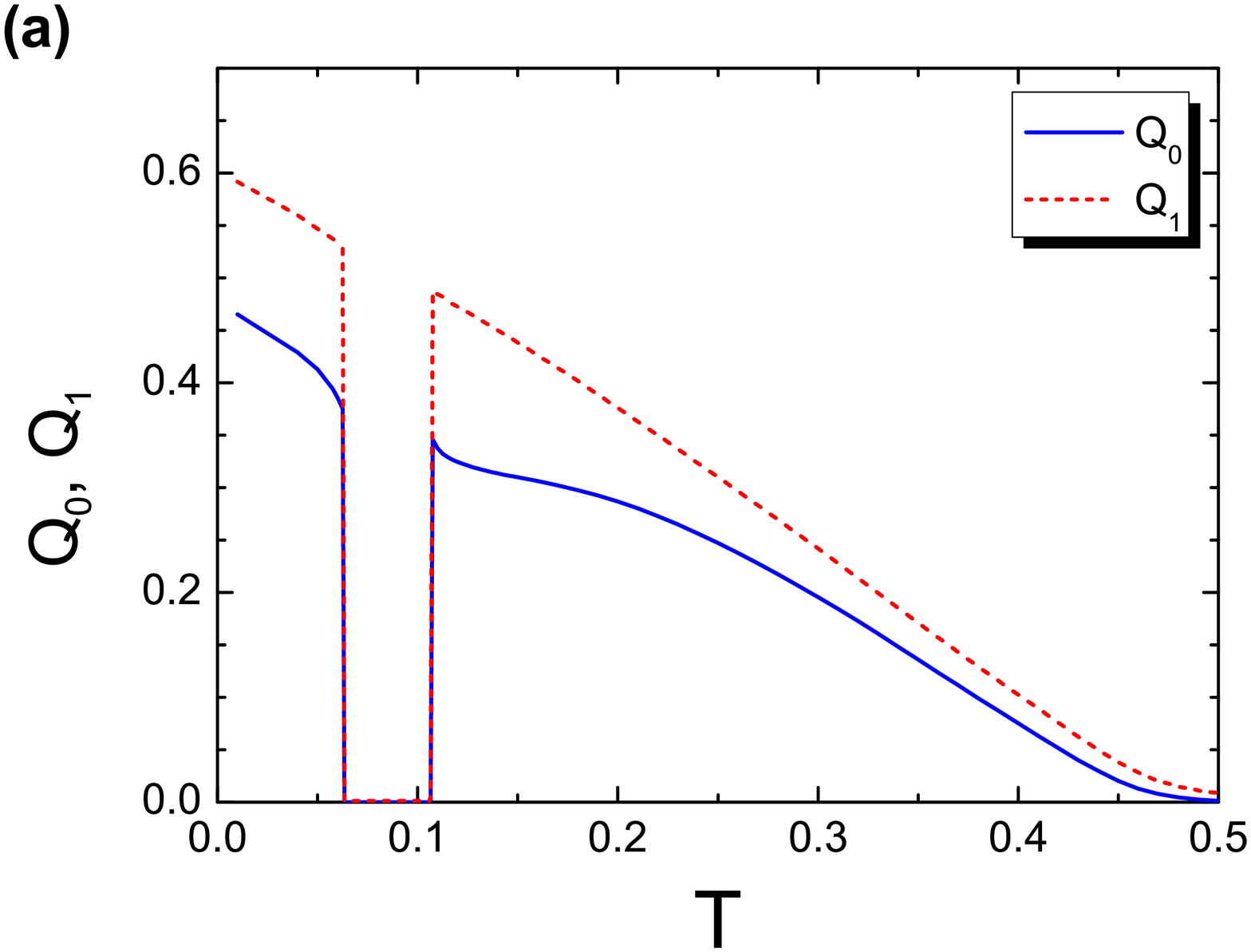}
\includegraphics[width=0.42\textwidth]{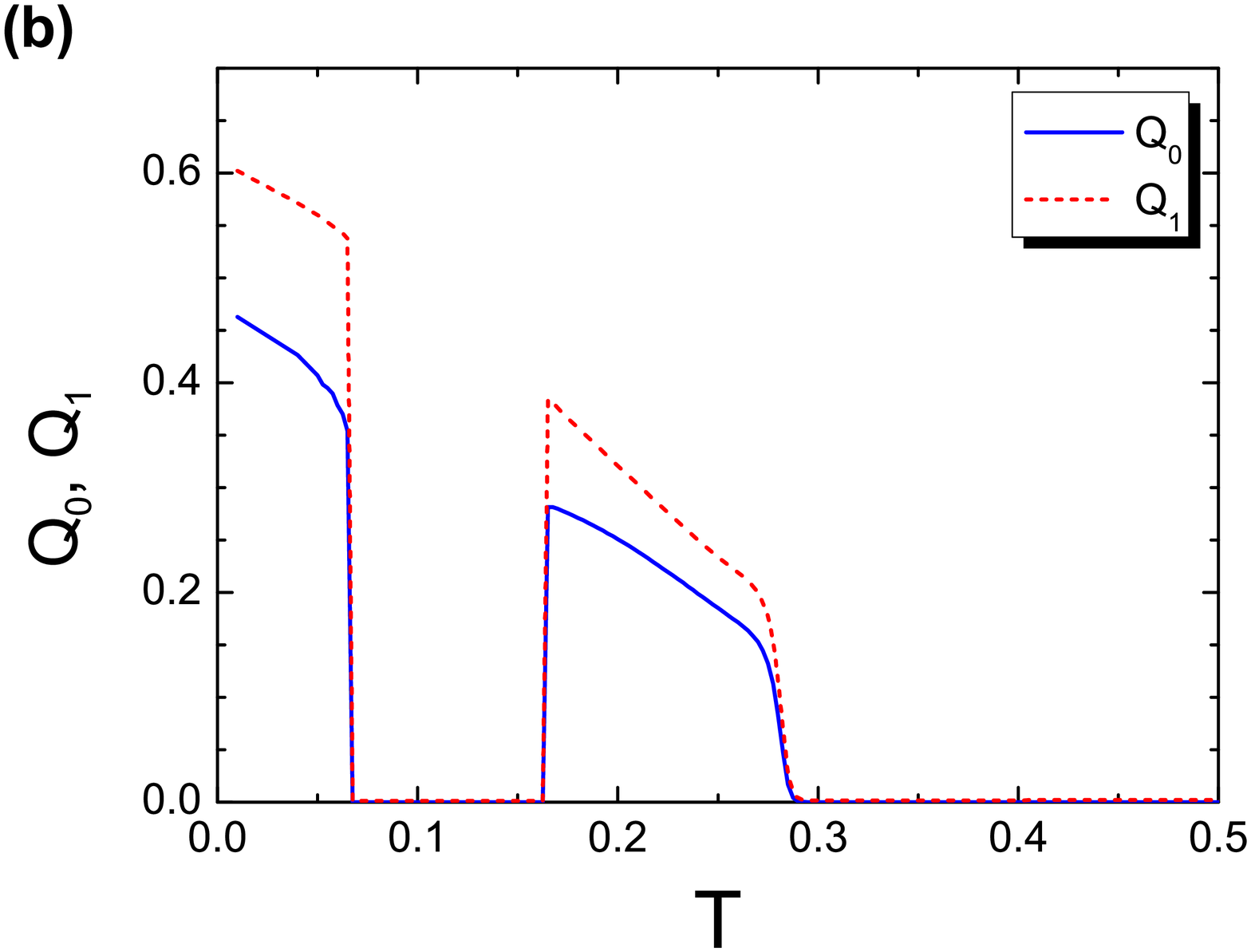}
\includegraphics[width=0.42\textwidth]{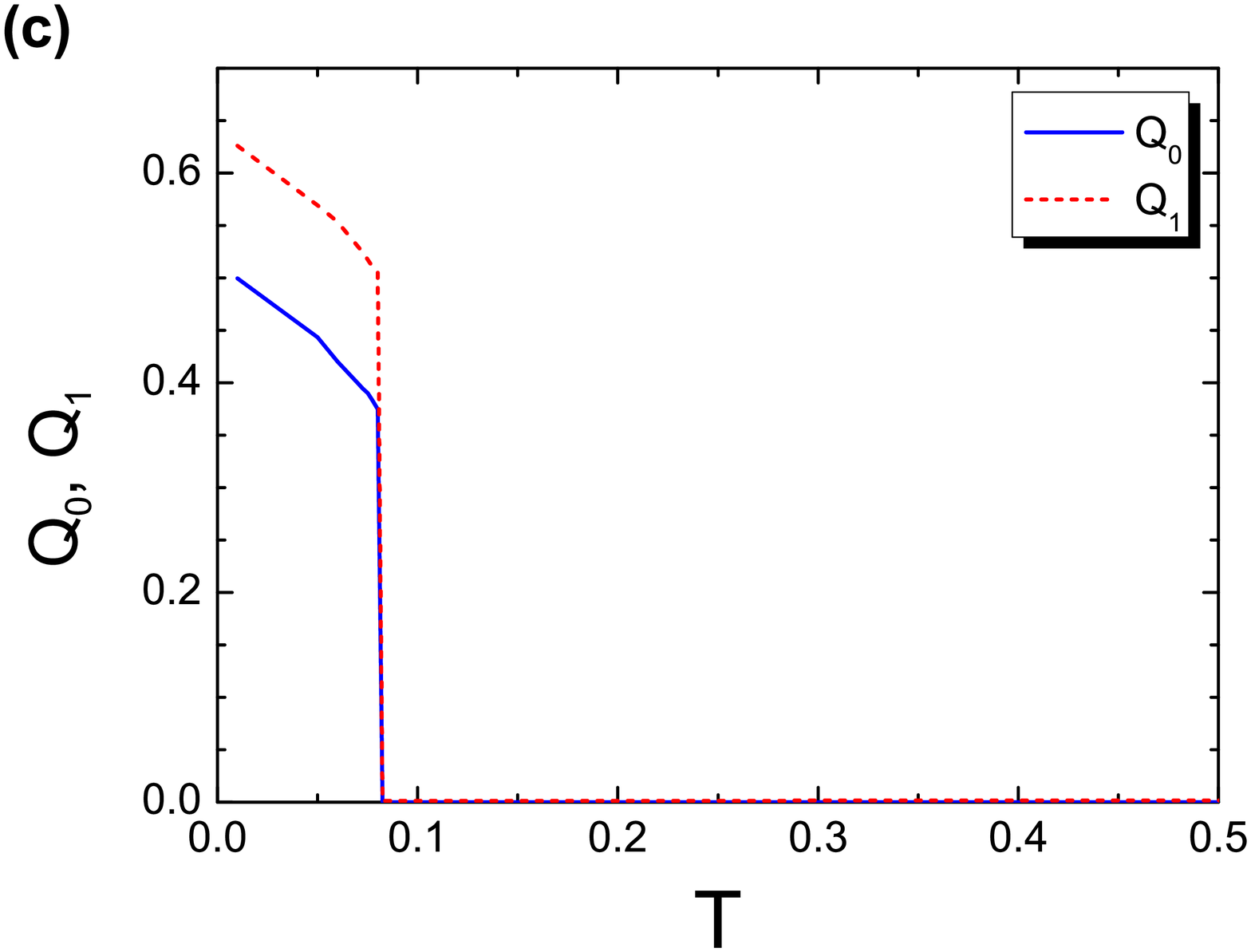}
\includegraphics[width=0.42\textwidth]{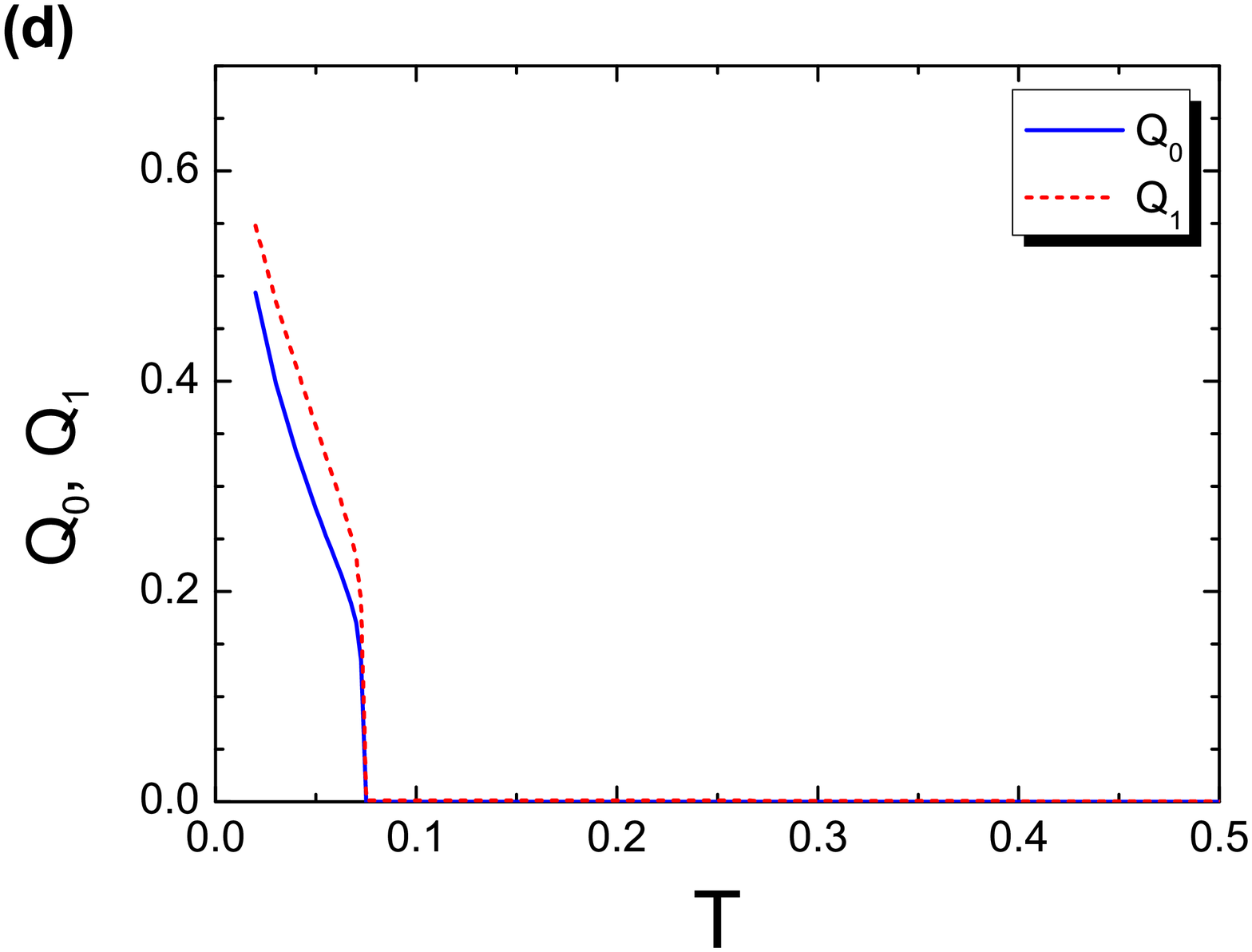}
\caption{(Color online) Values of $Q_{0}$ and $Q_{1}$ for (a) $\Gamma = 0.0$,
(b)	$\Gamma = 0.04$, (c) $\Gamma = 0.1$, and (d) $\Gamma = 0.55$. Here $D$ is fixed at 0.88.
}\label{3}
\end{figure}

Our previous results can be directly checked by numerical analysis of the free energy $f_{_{1RSB}}$, $Q_{0}$, and $Q_{1}$. All values of $Q_{0}$ and $Q_{1}$ shown in Fig. 3 are obtained for $D=0.88$, which is given for comparison with Fig. 2(c). For $\Gamma = 0.0$, as shown in Fig. 3(a), phase transitions occur in the order PM $\stackrel{\mathrm{2nd}}{\longrightarrow}$ HTSG $\stackrel{\mathrm{1st}}{\longrightarrow}$ PM $\stackrel{\mathrm{1st}}{\longrightarrow}$ LTSG as the temperature is reduced. Here the first-order phase transitions can be easily confirmed as sudden changes in the free energy $f_{_{1RSB}}$ or discontinuities of the entropy $S$, which is the temperature-derivative of the free energy $f_{_{1RSB}}$. The PM phase gap between HTSG and LTSG, i.e., the difference between the first-order transition temperature of the HTSG-to-PM transition and that of the PM-to-LTSG transition, is an extremely small value of 0.04. In Fig. 3(b), when $\Gamma$ is increased to 0.04, the PM phase gap between the HTSG and the LTSG widens to 0.1, and the phase transitions occur in the order PM $\stackrel{\mathrm{2nd}}{\longrightarrow}$ HTSG $\stackrel{\mathrm{1st}}{\longrightarrow}$ PM $\stackrel{\mathrm{1st}}{\longrightarrow}$ LTSG as the temperature is reduced. When $\Gamma$ is increased to 0.1, as shown in Fig. 3(c), the HTSG disappears and a first-order phase transition occurs from PM to LTSG as the temperature is decreased. This feature is maintained even when $\Gamma$ is increased to 0.55, which is shown in Fig. 3(d).

\begin{figure}
\includegraphics[width=0.5\textwidth]{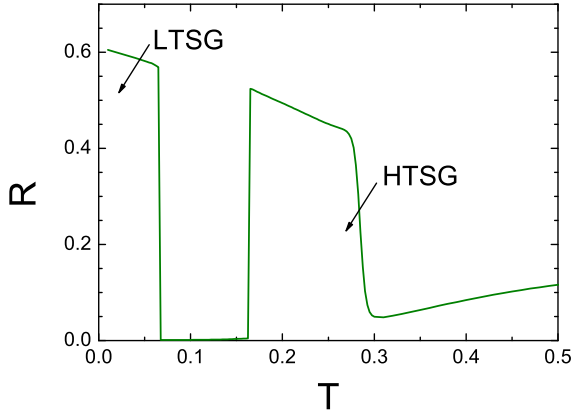}
\caption{(Color online) Values of $R$ for $\Gamma=0.04$ and $D=0.88$.}\label{4}
\end{figure}

\begin{figure}
\includegraphics[width=0.5\textwidth]{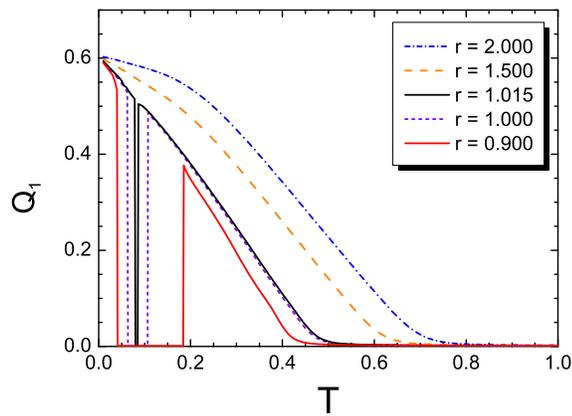}
\caption{(Color online) Values of $Q_{1}$ for $\Gamma=0.0$, $D=0.88$, and several $r$ values.}\label{5}
\end{figure}

In the inverse freezing among PM-SG-PM phases in the GS model, there has been a discovery that the higher-temperature PM phase is characterized by a low density of empty states, whereas the lower-temperature PM phase has a higher density of empty states \cite{Morais12}. Here the density of empty states $n_{0}$ plays a crucial role in distinguishing the two PM phases. Similarly, in order to clarify a difference between two SG phases, we draw a graph of the spin self-interaction $R$ [Eq.(10)], which signifies the density of filled states. As shown in Fig. 4, $R$ shows the difference between two SG phases clearly: The HTSG has lower $R$ values than the LTSG does. Thus, we can infer that the LTSG is characterized by a higher density of filled states.

We finally examine whether the second reentrance or the splitting between the HTSG and the LTSG occur at $r \neq 1$. As shown in Fig. 5, at $r=0.9$, the PM phase gap between the HTSG and the LTSG is wider than that of the $r=1$ case. At $r=1.015$, there is an extremely narrow gap near $T=0.082$. When $r$ is larger than 1.015, there exists only one SG phase, instead of the two separated SG phases. Since Schupper and Shnerb \cite{Schupper04} focused on the inverse freezing of the GS model, they selected large values of $r$ (e.g., 6.0). In order to observe SG splitting, however, it is better to select $r$ values smaller than 1.0 because when the degeneracy of the empty states of $S_{z}$ ($l$) is larger than one of the filled states of $S_{z}$ ($k$), the PM phase gap generating the SG splitting becomes wider.

\section{Conclusions}

In the present work, we proposed an expanded spin-glass model, the quantum GS model, in order to obtain more meaningful evidence for the second reentrance observed in the GS model. By obtaining the 1RSB solutions of the quantum GS model, we could check the detailed PM-SG phase boundaries depending on the crystal field $D$ and the transverse field $\Gamma$.
We first confirmed that a second reentrance occurs in the GS model ($\Gamma = 0.0$ case), as reported by Crisanti and Leuzzi \cite{Crisanti05,Leuzzi06}. We can thus describe the GS model as a prototypical model that can be used to verify the second entrance as well as inverse freezing.
Furthermore, there exist first-order phase transitions and TCPs for $\Gamma \geq 0.0$ and large values of $D$. This is clearly observable from the $T-\Gamma$ phase diagrams for $D \geq 0.7$, which are shown in Fig. 2(b). In particular, when $D$ is larger than 0.879, one SG phase is split into two SG phases (HTSG and LTSG). We can distinguish the two SG phases by the spin self-interaction $R$. The HTSG and LTSG show certain differences in shape and phase boundaries. Such SG splitting becomes more distinctive when $r$ is less than 1. We verified that the empty states of $S_{z}$ are thus crucial for the occurrence of SG splitting.

It is well known that the SK model with a transverse field \cite{Chakrabarti96,Kim02} has been successfully applied to the quantum spin glass
$\textrm{LiHo}_{x}\textrm{Y}_{1-x}\textrm{F}_{4}$ \cite{Wu}, a site-diluted and isostructural
derivative of the dipolar-coupled Ising ferromagnet $\textrm{LiHoF}_{4}$ ($T_{c}$=1.53K). In the absence of a
magnetic field, $\textrm{LiHo}_{x}\textrm{Y}_{1-x}\textrm{F}_{4}$ is a conventional spin glass with the glass transition temperature
$T_{g}(x)$. When an externally tunable magnetic field is induced transverse to the magnetic easy axis, quantum tunneling
occurs. Provided we can identify a suitable candidate spin-glass material with $S=1$ and crystal field and provided quantum tunneling by an externally tunable transverse magnetic field occurs in the material, we may be able to observe and verify SG splitting through experimental results.

In contrast, it would be of interest to extend our theory beyond the static approximation used in this work in order to obtain analytic solutions for free energy and order parameters. It would also be interesting theoretically to search for other SG models for SG splitting. We believe that these topics will extend our viewpoint on SG systems.

\begin{acknowledgments}

The author thanks Jesuit Community colleagues at the Sogang University for helpful comments. This work was supported by the Formation Fund for Korean Jesuit Scholastics.

\end{acknowledgments}

\end{document}